\documentclass[a4paper,aps,prd,10pt,preprintnumbers,showpacs,twocolumn,superscriptaddress,nofootinbib,amsmath,amssymb,floatfix]{revtex4-1}\def\JCAPstyle#1{}
\usepackage{enumitem}
\usepackage{graphicx}
\usepackage[utf8]{inputenc}
\usepackage[T1]{fontenc}
\usepackage{cmap}
\usepackage{subfigure}
\usepackage{color}
\usepackage[colorlinks=true,linkcolor=blue,citecolor=blue,urlcolor=blue]{hyperref}

\DeclareMathAlphabet{\pazocal}{OMS}{zplm}{m}{n}

\begin{document}

%\preprint{APS/-QNM}

\title{Thermodynamic topology of phantom AdS black holes in massive gravity}
\author{Hao Chen}% %
\email{haochen1249@yeah.net(Corresponding author)}
\affiliation{School of Physics and Electronic Science, Zunyi Normal University, Zunyi, Guizhou 563006, People's Republic of China}
\author{Di Wu}
\email{wdcwnu@163.com}
\affiliation{School of Physics and Astronomy, China West Normal University, Nanchong, Sichuan 637002, People's Republic of China}
\author{Meng-Yao Zhang}% %
\email{myzhang94@yeah.net}
\affiliation{College of Computer and Information Engineering, Guizhou University of Commerce, Guiyang, 550014, China}
\author{Hassan  Hassanabadi}% %
\email{h.hasanabadi@shahroodut.ac.ir}
\affiliation{Department of Physics, University of Hradec Kr\'{a}lov\'{e}, Rokitansk\'{e}ho 62, 500 03 Hradec Kr\'{a}lov\'{e}, Czechia }
\author{Zheng-Wen Long}% %
\email{zwlong@gzu.edu.cn }
\affiliation{College of Physics, Guizhou University, Guiyang, Guizhou 550025, People's Republic of China}

\date{\today }

\begin{abstract}
In this work, we explore the thermodynamic topology of phantom AdS black holes in the context of massive gravity. To this end, we evaluate these black holes in two distinct ensembles: the canonical and grand canonical ensembles (GCE). We begin by examining the topological charge linked to the critical point and confirming the existence of a conventional critical point $(CP_{1})$ in the canonical ensemble (CE), this critical point has a topological charge of $-1$ and acts as a point of phase annihilation, this situation can only be considered within the context of the classical Einstein-Maxwell (CEM) theory $(\eta=1)$, while no critical point is identified in the GCE. Furthermore, we consider black holes as a topological defect within the thermodynamic space. To gain an understanding of the local and global topological configuration of this defect, we will analyze its winding numbers, and observe that the total topological charge in the CE consistently remains at $1$. When the system experiences a pressure below the critical threshold, it gives rise to the occurrence of annihilation and generation points. The value of electric potential determines whether the total topological charge in the GCE is zero or one. As a result, we detect a point of generation point or absence of generation/annihilation point. Based on our analysis, it can be inferred that ensembles significantly impact the topological class of phantom AdS black holes in massive gravity.
\end{abstract}

\maketitle

\section{Introduction}

In 1974, Hawking made a groundbreaking contribution to the understanding of black holes by introducing the concept of Hawking radiation. This discovery not only resolved the inherent contradiction between black holes and thermodynamics but also established a profound correlation between gravity, quantum mechanics, and thermodynamics \cite{ch1}. When investigating the Schwarzschild black hole within the AdS framework, it is noted that the stability  hinges on a critical temperature known as the Hawking-Page (HP) phase transition \cite{ch2}. This HP phase transition holds significance within the AdS/CFT correspondence theory as it represents a transition between confinement and deconfinement in gauge/gravity duality \cite{ch3}. Kastor, Ray and Traschen initially proposed that the cosmological constant $\Lambda$ could serve as a representation of the thermodynamic pressure, implying that one could consider the conjugate quantity of the thermodynamical volume of the black hole in AdS space \cite{ch4,ch5,ch6,ch7}. This has generated considerable interest in investigating the thermodynamics \cite{CPL23-1096,PRD84-024037,PRD100-101501,PRD101-024057,PRD102-044007,PRD103-044014,JHEP1121031,PRD105-124013,PRD108-064034,PRD108-064035,
PLB846-138227,2406.13461}, phase transitions \cite{ch8,ch9,ch10,sc1,ch11,ch12,ch13,ch14,ch15,ch16,ch17,ch18,ch19,ch20,ch21,ch22}, and Joule-Thomson expansions \cite{ch29,ch30,ch31,ch32,ch33, ch34,ch35,ch36,ch37} of AdS black holes within the framework of the extended phase space.

Recently, the study of black hole criticality has involved the idea of thermodynamic topology, which can help reveal more about the nature of black holes. The Duan's topological current $\phi$-mapping theory \cite{kh6}, first introduced in Ref. \cite{kh5}, is utilized in this innovative method to examine the criticality of a black hole's thermodynamic space, the temperature function represents the distribution of topological charge to these critical points. On the basis of this assumption, the following can be done to determine the black hole's critical point
\begin{equation}\label{wei1}
\left(\partial_S T\right)_{P, x^i}=0, \quad\left(\partial_{S,S} T\right)_{P, x^i}=0.
\end{equation}
Derive a new temperature function from Eq.\ref{wei1} by eliminating thermodynamic pressure. Then, proceed to construct Duan's potential \cite{kh5}
\begin{equation}\label{ph2}
\Phi=\frac{1}{\sin \theta} T\left(S, x^i\right),
\end{equation}
 note that incorporating $1/\sin \theta$ introduces an additional factor that aids in facilitating topological analysis. To utilize Duan's $\phi$-mapping theory, a novel vector field denoted as $\phi=(\phi^S,\phi^\theta)$ is introduced
\begin{equation}
\phi ^S=(\partial _S \Psi )_{\theta ,x_i}, \phi ^\theta=(\partial _\theta \Psi )_{ S,x_i}.
\end{equation}
Motivated by work \cite{kh7}, when the value of $\theta$ is equal to $\frac{\pi}{2}$, the $\phi$ component of the vector field remains constant at zero. This permits the characterization of the topological current in the following manner
\begin{equation}
J^\mu=\frac{1}{2\pi } \epsilon ^{\mu \nu \lambda }\epsilon _{ab}\partial_{ \nu }n^a\partial_{ \lambda  }n^b,
\end{equation}
here, we define $\partial_{ \nu }$ as the derivative with respect to $x^{\nu}$, where $x^{\nu}$ represents the coordinates $(t, r, \theta)$. The unit vector $n$ is determined by its components $n^1$ and $n^2$, which are given by $n^1=\frac{\phi^S}{\left\| \phi \right\|}$ and $n^2=\frac{\phi^{\theta}}{\left\| \phi \right\|}$ respectively. The condition $\partial_\mu J^\mu=0$ must be satisfied in order to establish the existence of a topological current. Moreover, the topological charge associated with a parameter region $\sum $ can be mathematically represented as
\begin{equation}
Q_t=\int_{\sum} j^0 d^2 x=\sum_{i=1}^N w_i,
\end{equation}
The number of windings, denoted as $w_i$, corresponds to the locations where $\phi^{a}$ equals zero. The total topological charge, denoted as $ Q _ { t } $, can be calculated by aggregating the charges linked to individual critical points, which are influenced by their respective winding properties. The investigation of the thermodynamic topology of black holes has been expanded to encompass diverse categories of black holes \cite{o1,o2,o3,o4,o5,o6,o7,o8,o9,o10,o11,tp3,po1}.

A different method for integrating topology into the realm of black hole thermodynamics has been suggested in a recent study \cite{kh8}, this approach suggests that black hole solutions can be viewed as defects. The topological number, known as the total winding number, is utilized for classifying different solutions of black holes. The examination commences with the presentation of a generalized free energy, denoted as $F$, which is introduced by
\begin{equation}\label{wei8}
\mathcal{F}=E-\frac{S}{\tau},
\end{equation}
in this context, $E$ and $S$ denote energy and entropy correspondingly, and $\tau$ is a dimensionless quantity denoting time, the vector field $\phi$ is derived from
\begin{equation}\label{wei10}
\phi=\left(\frac{\partial \mathcal{F}}{\partial r_{+}},-\cot \Theta \csc \Theta\right).
\end{equation}
The vector $\phi$ has its zero point located at $\Theta=\pi / 2$, the unit vector is characterized by
\begin{equation}
n^a=\frac{\phi^a}{\|\phi\|} \quad(a=1,2) \quad \text { and } \quad \phi^1=\phi^{r_{+}}, \quad \phi^2=\phi^{\Theta}.
\end{equation}
The determination of the zero points of $n^1$ is contingent upon the given value of $\tau$, the computation of winding numbers is performed for each zero point. The total winding number can be derived by combining the separate winding numbers of each existing branch of black holes. The exploration as topological imperfections has expanded to encompass a wide range of diverse black holes \cite{l1,l2,l3,l4,l5,l7,l9,l10,l11,l12,l13,l14,l16,l17,l18,l19,l20,l21,l22}.

On the other hand, the study of massive gravity reveals how the graviton mass induces quantum corrections to black hole entropy and Hawking radiation, providing unique avenues to test these modifications. This approach offers novel insights into resolving the thermodynamical phase transitions and quantum fluctuations of black holes \cite{rew1,rew2,rew3,rew4,rew5,rew6}. Inspired by the aforementioned research, the objective of this paper is to broaden the exploration of thermodynamic topology in various ensembles while considering phantom AdS black holes in the massive gravity framework. To begin with, we conduct an individual examination of the ensembles to identify crucial points and assess their topological charges, these charges are classified as either conventional $(Q_{t}=-1)$ or novel  $(Q_{t}=1)$ critical points. Afterward,  black holes are evaluated as a topological defect, the topological number, annihilation, and generation points of the system are determined. The manuscript is organized in the following structure: In section \ref{sec2}, we will present a concise overview of the thermodynamic properties linked to phantom AdS black holes within the framework of massive gravity. Moving forward to section \ref{sec3}, our focus will be on investigating the thermodynamic topology of  black holes in the CE. Furthermore, we will expand our investigation to include the GCE in section \ref{sec4}. Finally, concluding remarks and observations can be found in section \ref{sec5}.

\section{Phantom AdS Black Holes in Massive Gravity}\label{sec2}

The modified gravity theory exhibits captivating properties that enable it to elucidate phenomena such as the acceleration of the  universe, dark matter, and dark energy, which are not accounted for by general relativity. In 1939, Pauli and Fierz first put forward a linear theory of massive gravity with mass gravitons \cite{mg1}. Vainshtein modifies this theory nevertheless, since it is incompatible with the solar system's observational evidence \cite{mg2}. This problem is nevertheless present in Boulware and Deser (BD) ghost field \cite{mg3,mg4}, which is resolved by introducing dRGT massive gravity \cite{mg5,mg7,mg8}. In this regard, the extensive exploration of various black hole solutions has been conducted under the framework of dRGT massive gravity \cite{mg9,mg10,mg11,mg12,mg13}.
The Lagrangian of dRGT massive gravity in d-dimensional form is provided by \cite{mg5,k2}
\begin{equation}\label{g1}
\mathcal{L}_{\text {massive }}=m_g^2 \sum_i^d c_i \mathcal{U}_i(g, f),
\end{equation}
here $m_g$ represents the parameter associated with graviton mass, while $c_i$ denotes arbitrary massive couplings. $\mathcal{U}_i$ are potentials for the self-interaction of gravitons that are constructed by
 \begin{equation}
\mathcal{K}^\mu{ }_\nu=\delta_\nu^\mu-\sqrt{g^{\mu \sigma} f_{a b} \partial_\sigma \phi^a \partial_\nu \phi^b},
 \end{equation}
 where, the metric function $g_{\mu \nu}$ and the auxiliary reference metric $f_{\mu \nu}$ are employed in the formulation of the mass pertaining to gravitons, while $\phi^a$ represents St\"uckelberg fields. In this case, $\mathcal{U}_i$ can be expressed as
\begin{equation}
\mathcal{U}_i=\sum_{y=1}^i(-1)^{y+1} \frac{(i-1) !}{(i-y) !} \mathcal{U}_{i-y}\left[\mathcal{K}^y\right],
\end{equation}
here $\mathcal{U}_{i-y}=1$ with $i=y$, the explicit representation of $\mathcal{U}_i$ can be formulated as follows:
\begin{equation}
\begin{aligned}
&\mathcal{U}_1=[\mathcal{K}], \\
&\mathcal{U}_2=[\mathcal{K}]^2-\left[\mathcal{K}^2\right],\\
&\mathcal{U}_3=[\mathcal{K}]^3-3[\mathcal{K}]\left[\mathcal{K}^2\right]+2\left[\mathcal{K}^3\right],\\
&\mathcal{U}_4=[\mathcal{K}]^4-6\left[\mathcal{K}^2\right][\mathcal{K}]^2+8\left[\mathcal{K}^3\right][\mathcal{K}]+3\left[\mathcal{K}^2\right]^2-6\left[\mathcal{K}^4\right].
\end{aligned}
\end{equation}
In the context of a 4-dimensional space-time framework, the action of dRGT massive gravity, in conjunction with a phantom Maxwell field with the cosmological constant\cite{kh1}, can be formulated as follows:
\begin{equation}\label{g2}
\mathcal{I}=\int \sqrt{-g}\left[\mathcal{R}-2 \Lambda-2 \eta F_{\mu \nu} F^{\mu \nu}-\mathcal{L}_{\text {massive }}\right] d^4 x,
\end{equation}
the $\mathcal{R}$ symbols are indicative of the Einstein-Hilbert action, and $\Lambda$ represent the scalar curvature and cosmological constant correspondingly, which can exhibit behavior as $\mathrm{dS}(\Lambda>0)$ or AdS $(\Lambda<0)$.  The third term corresponds to the interaction with the Maxwell field in the case of $\eta=1$, or a spin $-1$ phantom field  when $\eta=-1$ \cite{k3}. The final component in the aforementioned action (\ref{g1}) pertains to the Lagrangian associated with dRGT massive gravity. In this study, we employed geometric units. The Faraday tensor, with $A_\mu$ serving as the gauge potential, is formulated as
\begin{equation}
F_{\mu \nu}=\partial_\mu A_\nu-\partial_\nu A_\mu.
\end{equation}
By considering Eq. (\ref{g2}) and applying the variational principle, the field equations associated with both gravitational and gauge is given by
\begin{equation}
\begin{aligned}
G_{\mu \nu}-\Lambda g_{\mu \nu}-m_g^2 \chi_{\mu \nu} & =2 \eta\left(\frac{g_{\mu \nu}}{4} F_{\mu \nu} F^{\mu \nu}-F_{\mu \rho} F_\nu^\rho\right), \\
\partial_\mu\left(\sqrt{-g} F^{\mu \nu}\right) & =0
\end{aligned}
\end{equation}
with
\begin{equation}
\chi_{\mu \nu}=-\sum_{i=1}^{d-2} \frac{c_i}{2}\left[\mathcal{U}_i g_{\mu \nu}+\sum_{y=1}^i(-1)^y \frac{i !}{(i-y) !} \mathcal{U}_{i-y} \mathcal{K}_{\mu \nu}^y\right].
\end{equation}
If we assume that the gravitational mass is negligible ($m_{g}=0$), the resulting field equation remains fully consistent with the result given in Ref. \cite{kh1}. Taking this into consideration, Jafarzade et al. found that the non-zero term of $\mathcal{U}_i$ is only $\mathcal{U}_1$ and $\mathcal{U}_2$ considering the 4-dimensional space-time, while the quartic terms are all eliminated \cite{kh1}
\begin{equation}
\begin{aligned}
&\mathcal{U}_1=\frac{2 c}{r}, \\
&\mathcal{U}_2=\frac{2 c^2}{r^2},\\
&\mathcal{U}_i=0\quad where \quad i>2.
\end{aligned}
\end{equation}
On this basis, the metric is offered for stationary phantom charged black hole solutions in the context of massive gravity
\begin{equation}
d s^2=f(r) d t^2-\frac{d r^2}{f(r)}-r^2\left(d \theta^2+\sin ^2 \theta d \varphi^2\right),
\end{equation}
here the metric function reads
\begin{equation}\label{kk1}
f(r)=1-\frac{2 m_0}{r}-\frac{\Lambda r^2}{3}+\frac{\eta q^2}{r^2}+\frac{C_1 r}{2}+C_2
\end{equation}
with
\begin{equation}
\begin{aligned}
&C_1=m_g^2 c c_1, \\
&C_2=m_g^2 c^2 c_2.
\end{aligned}
\end{equation}
The integration constant $m_{0}$ is associated with the  total mass, while the presence of massive gravity effects gives rise to a constant term $C_{2}$ $(C_{2}<1)$. Furthermore, if the massive gravity effects is ignored \cite{k3,kh9}, metric function (\ref{kk1}) will become
\begin{equation}
f(r)=1-\frac{2 m_0}{r}-\frac{\Lambda r^2}{3}+\frac{\eta q^2}{r^2}.
\end{equation}
Furthermore, by employing the Ashtekar-Magnon-Das approach  \cite{kh2,kh3}, we can find
\begin{equation}\label{ll1}
M=\frac{m_0}{4 \pi}.
\end{equation}
In the expanded phase space, the thermodynamic pressure exhibits a strong correlation with the cosmological constant $\Lambda$, precisely expressed as $P = -\Lambda/(8\pi)$ \cite{ch4,PRD84-024037}.
Based on this premise, the determination of the thermodynamic quantities, such as the Hawking temperature, entropy, electric charge, and mass, can be achieved by the event horizon's radius $r_{+}$  \cite{kh1} (we use per unit volume $\mathcal{V}_2=1$)
\begin{equation}\label{che1}
T=\left.\frac{1}{4 \pi} \frac{d f(r)}{d r}\right|_{r=r_{+}}=\frac{1}{4 \pi r_{+}}+2 P r_{+}+\frac{C_1}{4 \pi}+\frac{C_2}{4 \pi r_{+}}-\frac{q^2 \eta}{4 \pi r_{+}^3},
\end{equation}
\begin{equation}\label{wei7}
S=\frac{\mathcal{A}}{4}
=\frac{\left.\int_0^{2 \pi} \int_0^\pi \sqrt{g_{\theta \theta} g_{\varphi \varphi}}\right|_{r=r_+}}{4}
=\frac{r_+^2}{4},
\end{equation}
\begin{equation}
Q=\frac{F_{t r}}{4 \pi} \int_0^{2 \pi} \int_0^\pi \sqrt{g} d \theta d \varphi=\frac{q}{4 \pi},
\end{equation}
\begin{equation}\label{wei6}
M=\frac{r_+^3 P}{3}+\frac{C_1 r_+^2}{16 \pi}+\frac{\left(1+C_2\right) r_+}{8 \pi}+\frac{\eta q^2}{8 \pi r_+}.
\end{equation}
It can be easily demonstrated that the thermodynamic quantities adhere to the first law of thermodynamics, as expressed
\begin{equation}\label{meng1}
d M=T d S+\eta U d Q+V d P+\mathcal{C}_1 d C_1+\mathcal{C}_2 d C_2.
\end{equation}
The conjugate quantities in Eq. \ref{meng1} reads
\begin{equation}\label{chc2}
T=\frac{\partial M}{\partial S},
\end{equation}
\begin{equation}
\eta U  =\left(\frac{\partial M}{\partial Q}\right)_{S, P, C_i},
\end{equation}
\begin{equation}
V=\left(\frac{\partial M}{\partial P}\right)_{S, Q, C_i}=\frac{r_+^3}{3},
\end{equation}
\begin{equation}
\mathcal{C}_1 =\left(\frac{\partial M}{\partial C_1}\right)_{S, Q, C_2}=\frac{r_+^2}{16 \pi},
\end{equation}
\begin{equation}
\mathcal{C}_2  =\left(\frac{\partial M}{\partial C_2}\right)_{S, Q, C_1}=\frac{r_+}{8 \pi}.
\end{equation}
By utilizing the enthalpy mentioned in Eq. (\ref{wei6}) as a foundational reference, it becomes feasible to test the Hawking temperature using the established thermodynamic formula presented in Eq. (\ref{chc2}). In this case, the Smarr relation reads
\begin{equation}
M=2(T S-P V)+\eta U Q-\mathcal{C}_1 C_1.
\end{equation}
In the Smarr relation, $C_2$ is not included as it possesses a scaling weight of 0 \cite{hendi}, noted that the $C_2$ in the metric function represents a constant value in a 4-dimensional space time and does not contribute to thermodynamics.

\section{Phantom AdS Black Holes in Massive Gravity in  CE }\label{sec3}

In this section, we will delve into exploring the topology of phantom AdS black holes in the massive gravity thermodynamics. Specifically, we consider the temperature in Eq. (\ref{che1}) as a function
 \begin{equation}\label{wei2}
T=-\frac{q^2 \eta}{4 \pi r_{+}^3}+\frac{1}{4 \pi r_{+}}+2 P r_{+}+\frac{C_1}{4 \pi}+\frac{C_2}{4 \pi r_{+}}.
\end{equation}
By utilizing the condition described in Eq. (\ref{wei1}), we arrive at an expression for pressure
\begin{equation}
P=-\frac{3 q^2 \eta}{8 \pi r_{+}^4}+\frac{1}{8 \pi r_{+}^2}+\frac{C_2}{8 \pi r_{+}^2},
\end{equation}
then substitute the pressure $P$ into Eq. (\ref{wei2}), and eliminate the pressure term, yielding the following expression for the temperature $T$
\begin{equation}\label{wei3}
T=\frac{C_2}{2 \pi r_{+}}-\frac{q^2 \eta}{\pi r_{+}^3}+\frac{1}{2 \pi r_{+}}+\frac{C_1}{4 \pi}.
\end{equation}
A thermodynamic function, labeled as $\Phi$, is established by
\begin{equation}
\Phi=\frac{1}{\sin (\theta)}\left(\frac{C_2}{2 \pi r_{+}}-\frac{q^2 \eta}{\pi r_{+}^3}+\frac{1}{2 \pi r_{+}}+\frac{C_1}{4 \pi}\right),
\end{equation}
 the vector field $\phi=\left(\phi^{r_{+}}, \phi^\theta\right)$ encompasses the following vector components:
\begin{equation}
\phi^{r_{+}}=-\frac{\operatorname{csc}\theta\left(-6 q^2 \eta+r_{+}^2\left(1+C_2\right)\right)}{2 \pi r_{+}^4},
\end{equation}
and
\begin{equation}
\phi^\theta=-\frac{\cot \theta \csc \theta\left(-4 q^2 \eta+r_{+}^2\left(2+r_{+} C_1+2 C_2\right)\right)}{4 \pi r_{+}^3}.
\end{equation}
\begin{figure}
		\centering
        \includegraphics[scale = 0.48]{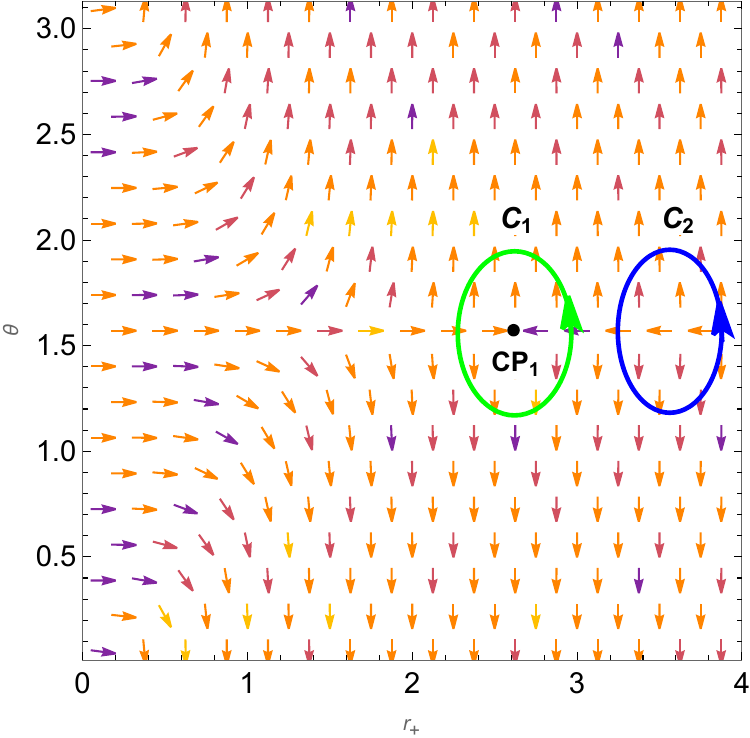} \hspace{-0.5cm}
\caption{The normalized vector field n in the $r_{+}$ vs $\theta$ plane is shown with a critical point indicated by a black dot.}
		\label{fig1}
	\end{figure}
Figure \ref{fig1} depicts the normalized vector $n$ and its vector plot for considered black holes, shown within the $r_{+}$ versus $\theta$ plane. Parameters for this plot include $q=\eta=1$ and $C_{1}=C_{2}=0.1$. The critical point, symbolized by the black dot ($CP_1$), is positioned at coordinate $(2.58199,
\pi/2)$.

\begin{figure}
		\centering
		\includegraphics[scale = 0.35]{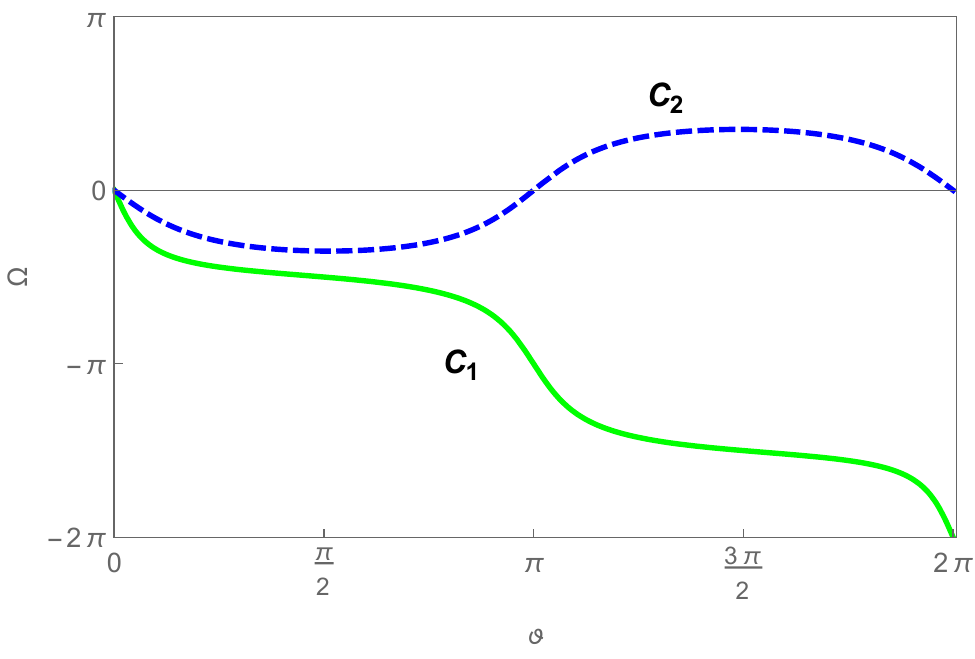} \hspace{-0.5cm}
\caption{$\Omega$ and $\vartheta$ is presented for contour $C1$, with parameters $a=0.15$, $b=0.4$, and $r_{0}=2.581989$, as well as for contour $C2$ with parameters $a=0.15$, $b=0.4$, and $r_{0}=3.6$.}
		\label{fig2}
	\end{figure}

To ascertain the topological charge associated with the critical point, the contour $C$ introduced in Ref. \cite{kh5} is defined by the parameterization $(\vartheta \in$ $0,2 \pi)$
\begin{equation}
\begin{aligned}
& r_{+}=a \cos \vartheta+r_0, \\
& \theta=b \sin \vartheta+\frac{\pi}{2}.
\end{aligned}
\end{equation}
The contour $C$ is used to determine the deflection angle $\Omega (\vartheta )$ of the vector field $\phi$, this angle is defined as follows:
\begin{equation}
\Omega(\vartheta)=\int_0^{\vartheta} \epsilon_{a b} n^a \partial_{\vartheta} n^b d \vartheta,
\end{equation}
the topological charge can be computed by
\begin{equation}
Q=\frac{\Omega (2\pi)}{2\pi}.
\end{equation}
Taking this into consideration, at the critical point denoted as $CP_1$ and enclosed by contour $C_1$, a conventional critical point is indicated with a topological charge value of $-1$. The contour $C_2$ results in a topological charge of zero due to the fact that doesn't contain any critical point. Hence, the total topological charge amounts to $Q=-1$, and the graphical representation of $\Omega$ can be observed in Fig. \ref{fig2}.

Furthermore, the equation of state (\ref{wei3}) allows us to discover the critical quantity as follows:
\begin{equation}\label{wei4}
\begin{aligned}
& r_{c}=\frac{q \sqrt{6 \eta}}{\sqrt{1+C_2}}, \\
& T_{c}=\frac{3 \sqrt{6 \eta} q C_1+4\left(1+C_2\right)^{3 / 2}}{12 \pi q \sqrt{6 \eta}}, \\
& P_{c}=\frac{\left(1+C_2\right)^2}{96 \pi q^2 \eta}.
\end{aligned}
\end{equation}
The result obtained in Ref. \cite{kh9} aligns perfectly with the critical point ($r_c=\sqrt{6} q, T_c=\frac{1}{3 \sqrt{6} \pi q}, \quad P_c=\frac{1}{96 \pi q^2}$) when the Massive Gravity parameters $C_{1},C_{2}$ are zero, and the theory is assumed to be Einstein-Maxwell-AdS black hole ($\eta=1$).
\begin{figure}
\centering
\includegraphics[scale = 0.35]{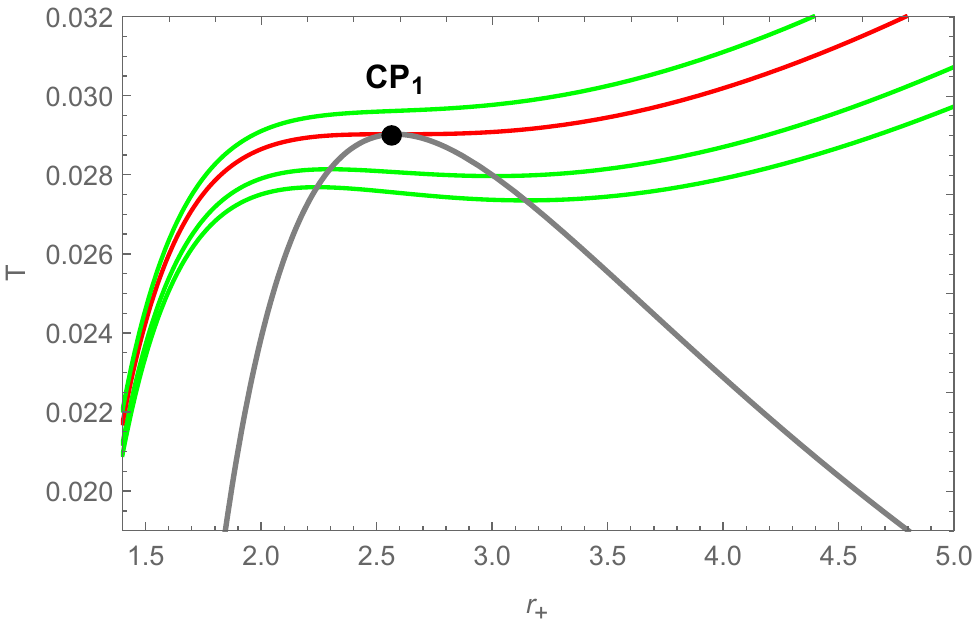} \hspace{-0.5cm}
\caption{Represented by a black dot, the critical point is depicted on the isobaric curves of black holes.}
\label{fig3}
\end{figure}

In Fig. \ref{fig3}, we depict the isobaric curves surrounding the critical point, revealing its nature. A black dot indicates the location of the critical point on each isobaric curve. The red curve represents the $P=P_c$ isobaric curve. The green curves are positioned for $P>P_c$ and $P<P_c$. By plotting the gray curve using Eq. (\ref{wei4}), we describe the extremal spots. The unstable zone divides the small and giant black hole phases for $P<P_c$, as shown in Fig. \ref{fig3}. At the critical point, various phases of black holes in the canonical ensemble converge. Thus, one way to define $CP_1$ as a phase annihilation point.

Now, we will investigate the phantom AdS black holes as topological thermodynamic defects in massive gravity. By employing the mass and entropy in Eqs. (\ref{wei7}) and (\ref{wei6}) in Eq. (\ref{wei8}), the generalized free energy is established by
\begin{equation}
\mathcal{F}=\frac{\mathrm{q}^2 \eta}{8 \pi r_{+}}+\frac{r_{+}}{8 \pi}-\frac{r_{+}^2}{4 \tau}+\frac{\mathrm{Pr}_{+}^3}{3}+\frac{r_{+}^2 C_1}{16 \pi}+\frac{r_{+} C_2}{8 \pi},
\end{equation}
the components of the vector field represented in Eq. (\ref{wei10}) are
\begin{equation}
\phi^{r}=\frac{1}{8 \pi}-\frac{q^2 \eta}{8 \pi r_{+}^2}-\frac{r_{+}}{2 \tau}+P r_{+}^2+\frac{r_{+} C_1}{8 \pi}+\frac{C_2}{8 \pi},
\end{equation}
and
\begin{equation}
\phi^{\Theta}=-\cot \Theta \csc \Theta.
\end{equation}
\begin{figure}
\centering
\includegraphics[scale = 0.28]{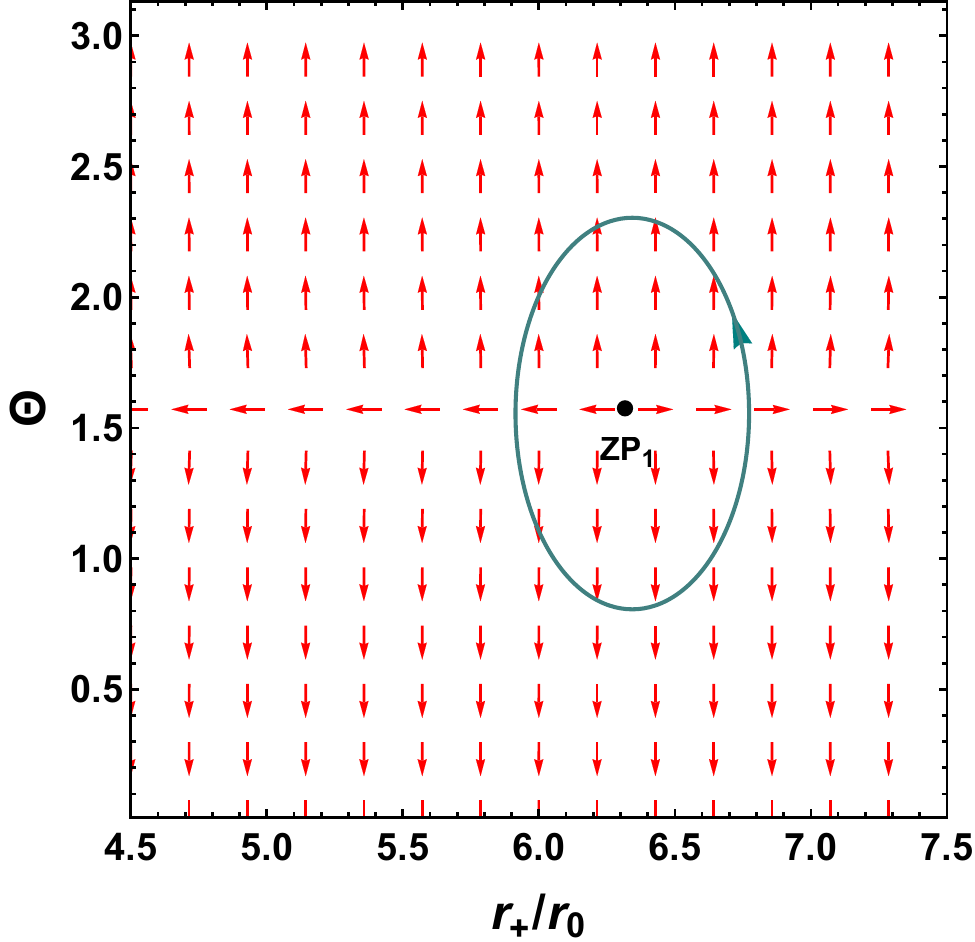} \hspace{-0.5cm}
\includegraphics[scale = 0.28]{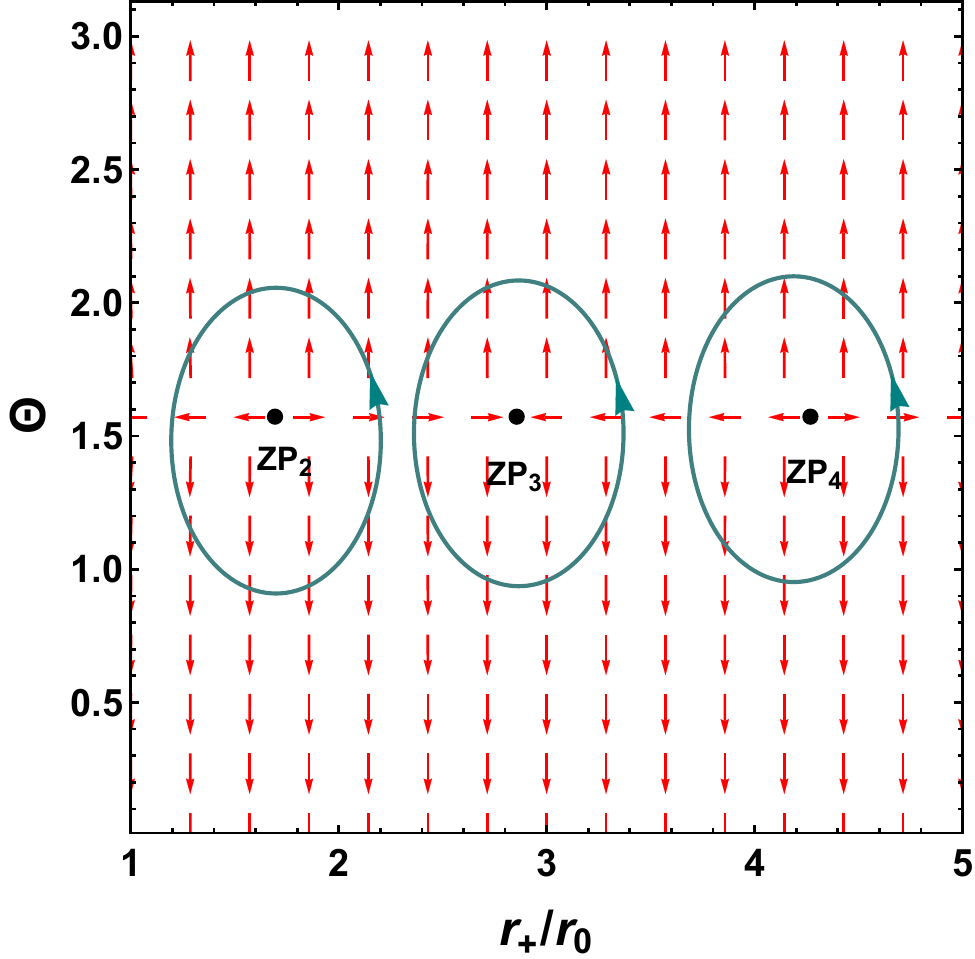} \hspace{-0.5cm}
\includegraphics[scale = 0.28]{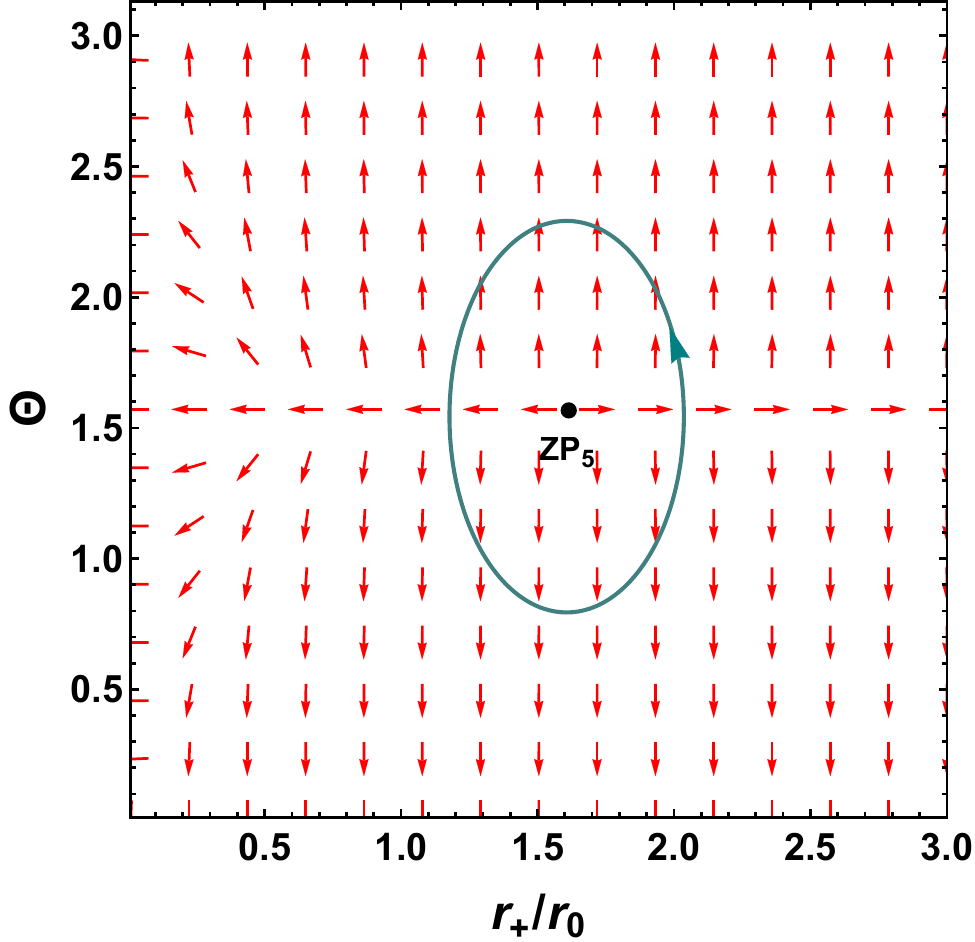} \hspace{-0.5cm}
\caption{Unit vector $n=\left(n^1, n^2\right)$ shown in $\Theta$ vs $r_{+} / r_0$ plane for $P r_0^2=0.0002$ (below critical pressure
$\left.P_c=0.002686\right)$. The black dots represent the zero points $(q/r_0=\eta/r_0=1,C_{1}/r_0=-0.3,C_{2}/r_0=-0.1)$. \textbf{Up panel:}$(\tau / r_0=80)$,\textbf{Middle panel:}$(\tau / r_0=110)$,\textbf{ Down panel:}$(\tau / r_0=130)$}
\label{fig6}
\end{figure}
\begin{figure}
\centering
\includegraphics[scale = 0.35]{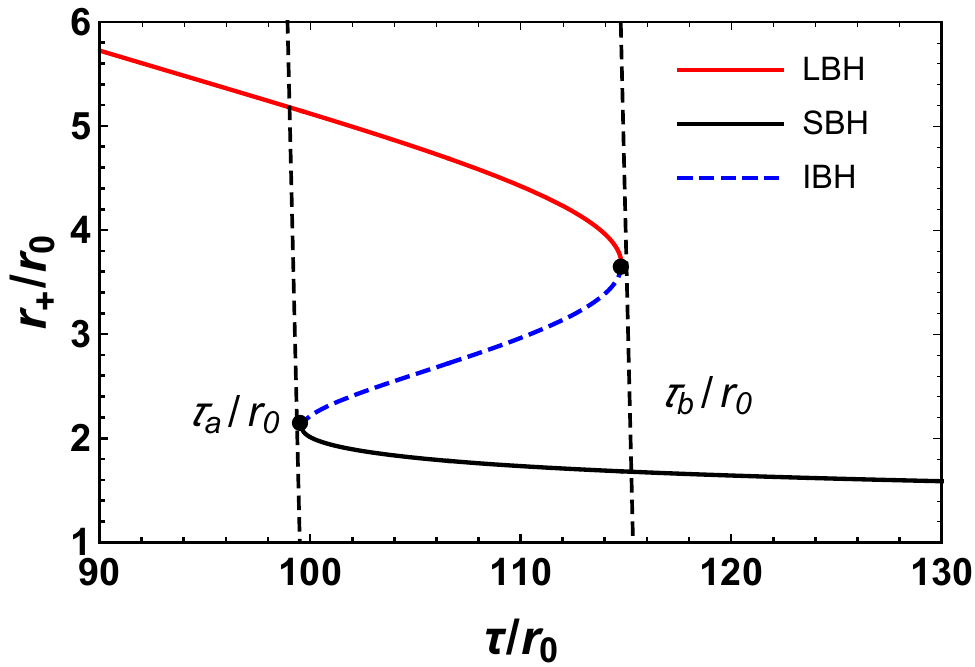} \hspace{-0.5cm}
\caption{The zero points of $\phi^{r+}$ in $\tau / r_0$ vs $r_{+} / r_0$ plane for $P r_0^2=0.0002$ (below critical pressure)$(q/r_0=\eta/r_0=1,C_{1}/r_0=-0.3,C_{2}/r_0=-0.1)$.}
\label{fig11}
\end{figure}
In Fig. \ref{fig6}, by setting $\tau/r_0=80$, we pinpoint a zero point $\left(ZP_1\right)$ located at $\left(r_{+}/r_0, \Theta\right)=(6.3523, \pi/2)$.  In this context, $r_0$ represents a length scale that is determined by the dimensions of the cavity surrounding the black hole, and can be chosen arbitrarily. By considering a pressure value below the critical pressure $P_c$, we confirm that the topological charge associated with this zero point is determined using the methodology outlined in the preceding section, resulting in $w=+1$. Similarly, We can also observe three points of zero: $ZP_2 (1.7322, \pi/2)$, $ZP_3 (2.96309, \pi/2)$, and $ZP_4 (4.42224, \pi/2)$, these points correspond to winding numbers of $+1$, $-1$, and $+1$ respectively when the ratio $\tau/r_0$ is set at $110$. For $\tau/r_0=130$, a positive winding number of $+1$ is observed in Fig. (\ref{fig6}), where a zero point denoted as $ZP_5$ can be found.

By finding the solution to $\phi^{r}=0$, we can find
\begin{equation}
\tau = \frac{4 \pi r_{+}^3}{-q^2 \eta+r_{+}^2+8 P \pi r_{+}^4+r_{+}^3 c_1+r_{+}^2 c_2}.
\end{equation}
It is possible to easily differentiate between three separate categories of black holes in Fig. (\ref{fig11}), each representing different regions within the parameter space. The branch $\tau<\tau_b$ corresponds to the region where large black holes are found, the zero points all have a consistent winding number of $w=+1$. Similarly, the branch $\tau>\tau_a$ characterizes the region of small black holes, also exhibiting a winding number of $w=+1$ for any zero point. The range $\tau_a<\tau<\tau_b$ represents the intermediate black hole region, which is distinguished by a winding number of $w=-1$ for every zero point situated within this branch, therefore, the topological number is determined to be $+1$ by performing the calculation of $W=+1-1+1$. It's important to highlight that the branch exhibiting a winding number of $+1$ signifies thermodynamic stability, as evidenced by a positive heat capacity. Conversely, the branch characterized by a winding number of $-1$ suggests thermodynamic instability, as indicated by a negative heat capacity.
\begin{figure}
		\centering
        \includegraphics[scale = 0.3]{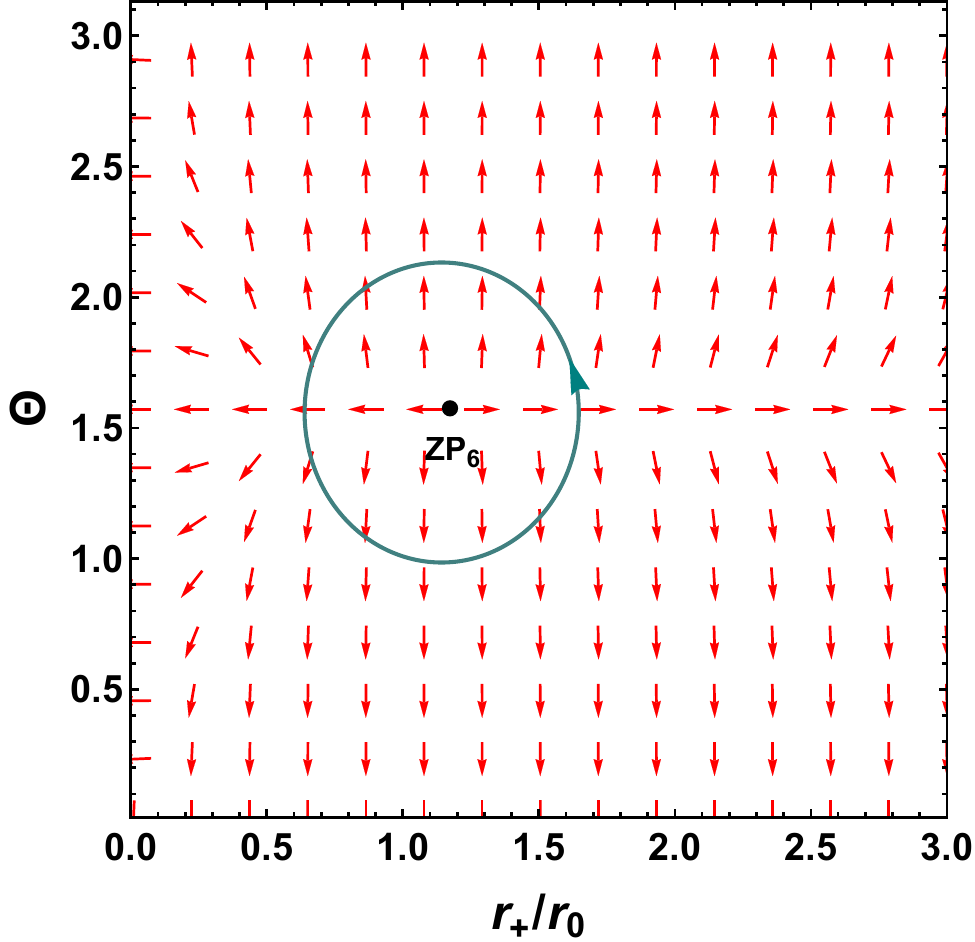} \hspace{-0.5cm}
		\includegraphics[scale = 0.3]{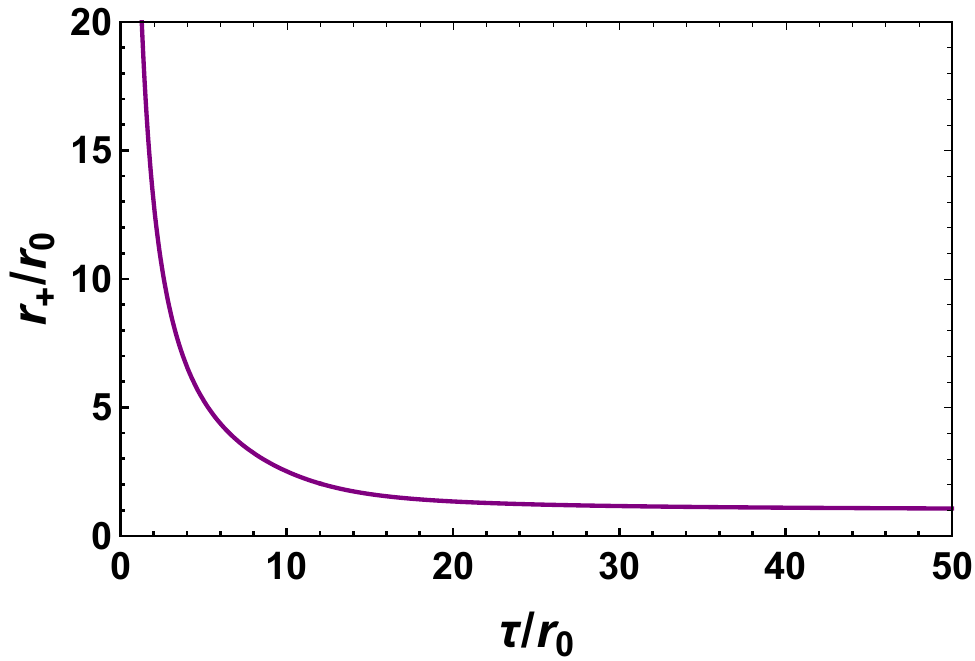} \hspace{-0.5cm}
\caption{\textbf{Up panel:} Unit vector $n=\left(n^1, n^2\right)$ shown in $\Theta$ vs $r_{+} / r_0$ plane for $\operatorname{Pr}_0^2=0.02$. The black dot represents zero point $(q/r_0=\eta/r_0=1,C_{1}/r_0=-0.3,C_{2}/r_0=-0.1)$.\\\textbf{ Down panel:} The zero points of $\phi^{r+}$ in $\tau / r_0$ vs $r_{+} / r_0$ plane for black holes in CE for pressure greater than the critical pressure $P_c$. $(q=\eta/r_0=1,C_{1}/r_0=-0.3,C_{2}/r_0=-0.1)$}
		\label{fi1}
	\end{figure}
Furthermore, we pinpoint the generation point at $\tau / r_0=\tau_a / r_0=99.6143$ and the annihilation point at $\tau / r_0=\tau_b / r_0=114.787$, respectively, represented by black dots in Fig. \ref{fig11}.

In the event where the pressure ($Pr_{0}^{2}=0.02$) exceeds the critical pressure in Fig. \ref{fi1}, we can observe that there is only one branch available, which represents a stable black hole exhibiting positive specific heat. The calculation of the winding numbers for zero points on this particular branch yields $w=+1$, leading to a topological number of $W=+1$. Noted that no generation/annihilation point has been detected under these circumstances. Furthermore, after varying both $q=\eta/r_0$ and the massive gravity parameters $C_{1}/r_0$ and $C_{2}/r_0$, we also find the topological number of  $W=+1$, this suggests that any changes made to the charge configuration will not affect the topological number of black holes.

\section{Phantom AdS Black Holes in Massive Gravity in  the GCE}\label{sec4}
\begin{figure}
		\centering
        \includegraphics[scale = 0.45]{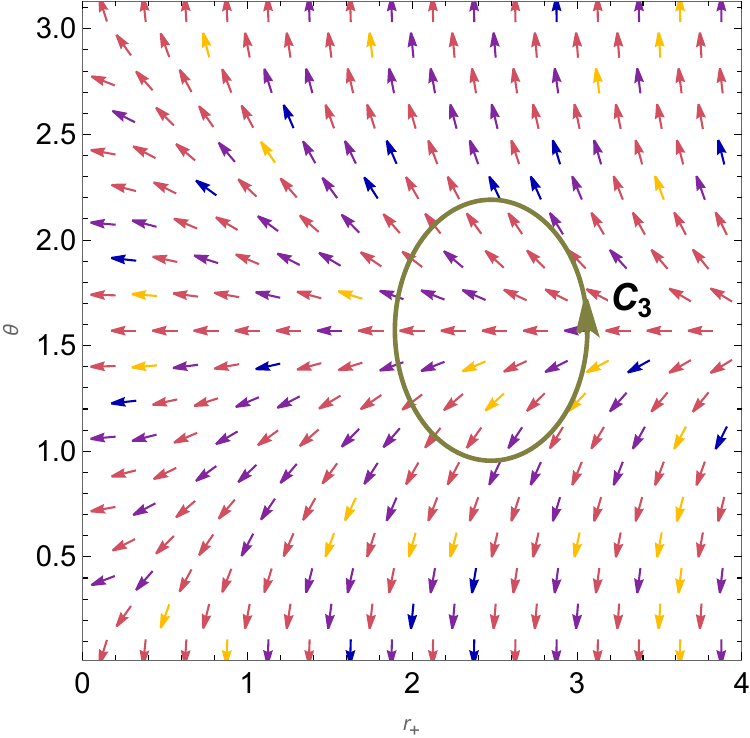} \hspace{-0.5cm}
		\includegraphics[scale = 0.4]{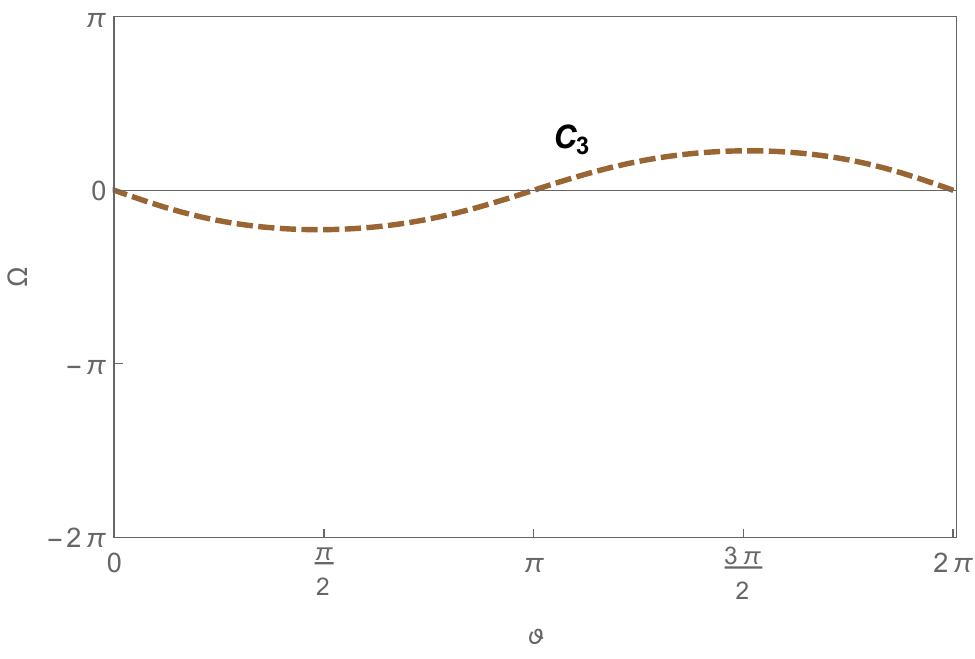} \hspace{-0.5cm}
\caption{\textbf{Up panel:} The normalized vector field n in the $r_{+}$ vs $\theta$ plane is shown without critical point indicated by a black dot $(U=0.1,C_{1}=-0.1, C_{2}=-0.3, \eta=-1)$. \textbf{ Down panel:} $\Omega$ and $\vartheta$ is presented for contour $C3$, with parameters $(U=0.1,C_{1}=-0.1, C_{2}=-0.3, \eta=-1, a=0.15,b=0.4,r_{0}=2.5)$}
		\label{f1}
	\end{figure}
\begin{figure}
		\centering
        \includegraphics[scale = 0.3]{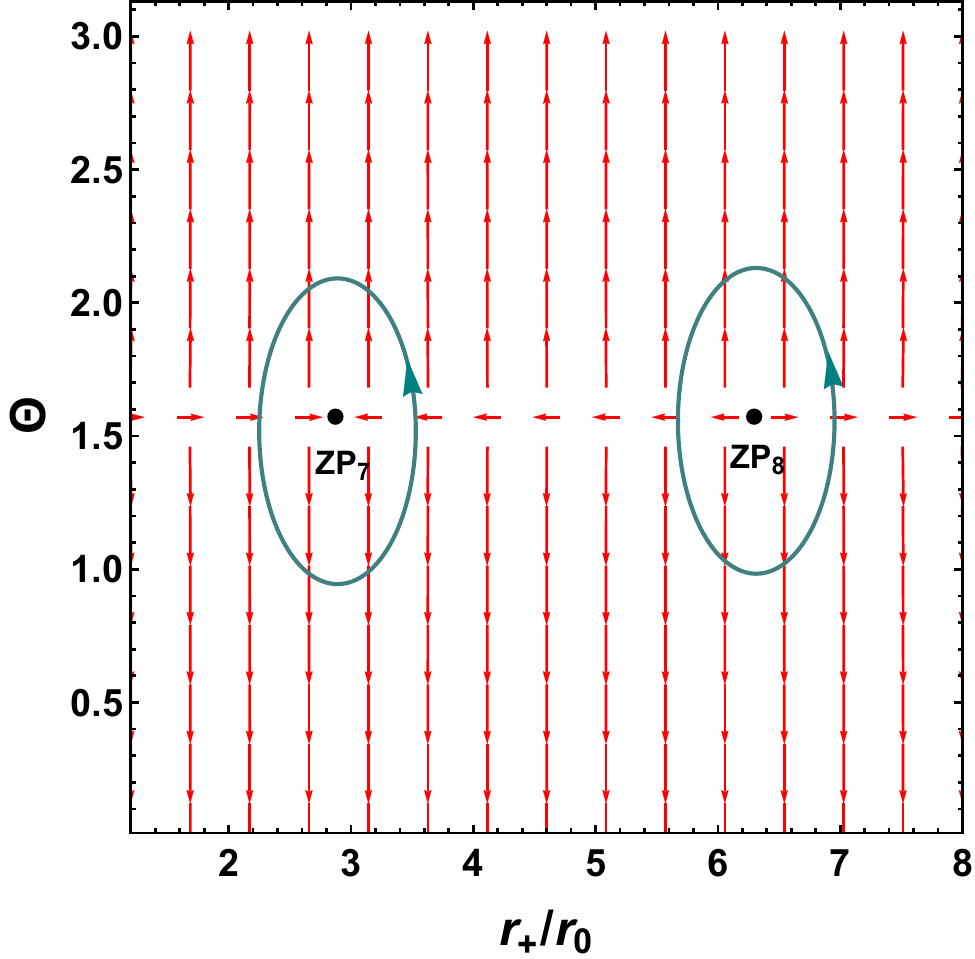} \hspace{-0.5cm}
		\includegraphics[scale = 0.3]{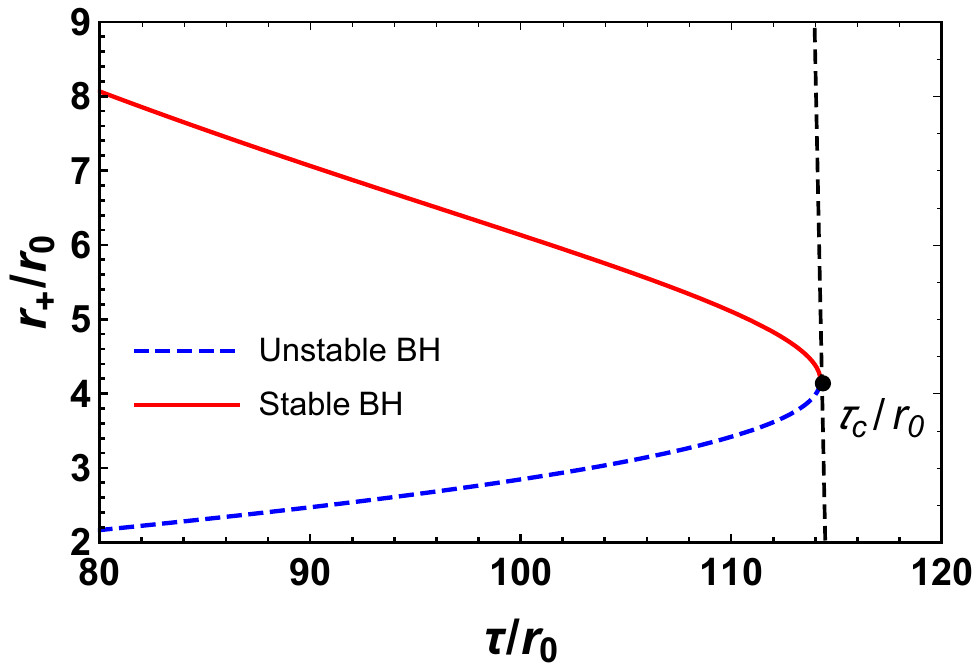} \hspace{-0.5cm}
\caption{ \textbf{Up panel:}Unit vector $n=\left(n^1, n^2\right)$ shown in $\Theta$ vs $r_{+} / r_0$ plane for $\tau / r_0=100, U/ r_0=0.1, C_{1}/ r_0=-0.1,C_{2}/ r_0=-0.3$ and $P r_0^2=0.001$. The black dots represent the zero points.\textbf{ Down panel:}The zero points of $\phi^{r+}$ in $\tau / r_0$ vs $r_{+} / r_0$ plane for black holes in GCE ($ U/ r_0=0.1, C_{1}/ r_0=-0.1,C_{2}/ r_0=-0.3$ and $P r_0^2=0.001$)}
		\label{nn1}
	\end{figure}
In this section, we will delve into exploring black holes in GCE within the context of massive gravity thermodynamics. Specifically, the electric  potential $U$ remains invariant \cite{kh1}, and can be expressed as
\begin{equation}
U=\frac{q}{r_+}.
\end{equation}
Following the methodology outlined in the preceding section, in this ensemble, the thermodynamic quantities are given by
\begin{equation}\label{pp1}
M =\frac{r_{+}}{8 \pi}+\frac{U^2 \eta r_{+}}{8 \pi}+\frac{P r_{+}^3}{3}+\frac{r_{+}^2 C_1}{16 \pi}+\frac{r_{+} C_2}{8 \pi},
\end{equation}
\begin{equation}
S=\frac{r_e^2}{4},
\end{equation}
\begin{equation}\label{ph1}
T = \frac{1}{4 \pi r_{+}}+\frac{U^2 \eta}{4 \pi r_{+}}+2 P_{r_{+}}+\frac{C_1}{4 \pi}+\frac{C_2}{4 \pi r_{+}}.
\end{equation}
\begin{figure}
		\centering
        \includegraphics[scale = 0.3]{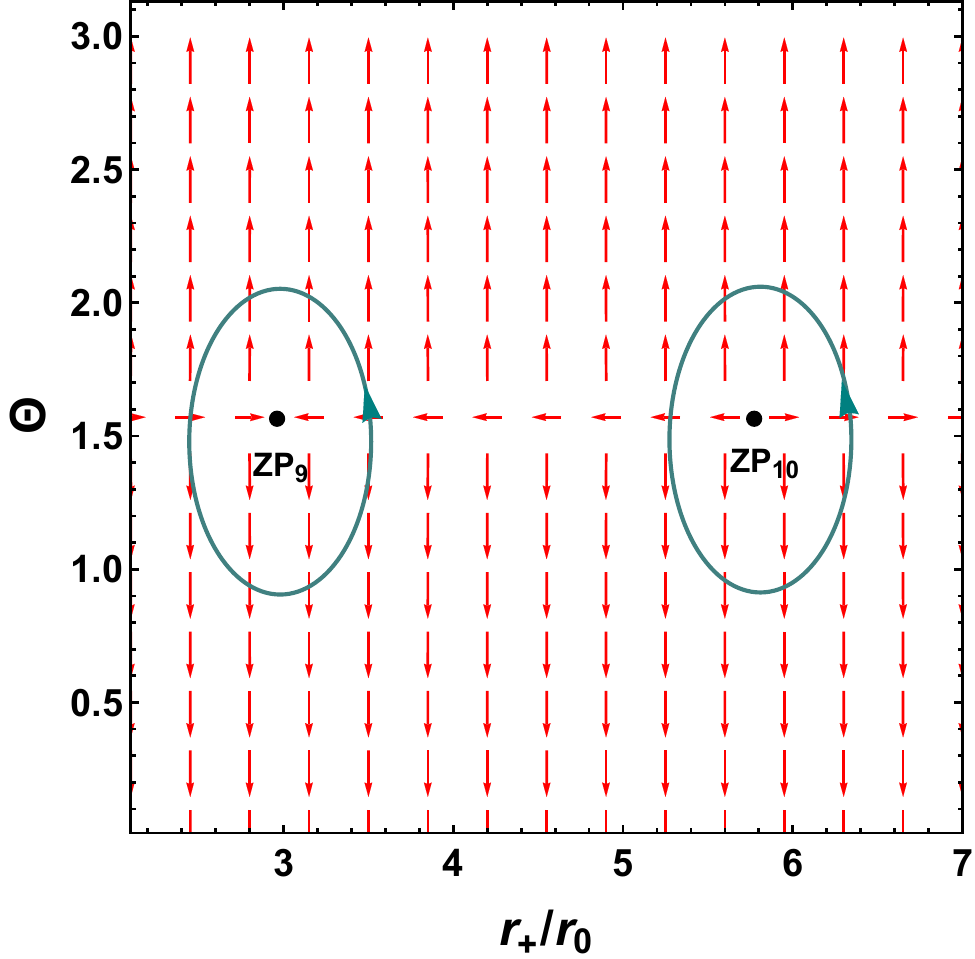} \hspace{-0.5cm}
		\includegraphics[scale = 0.3]{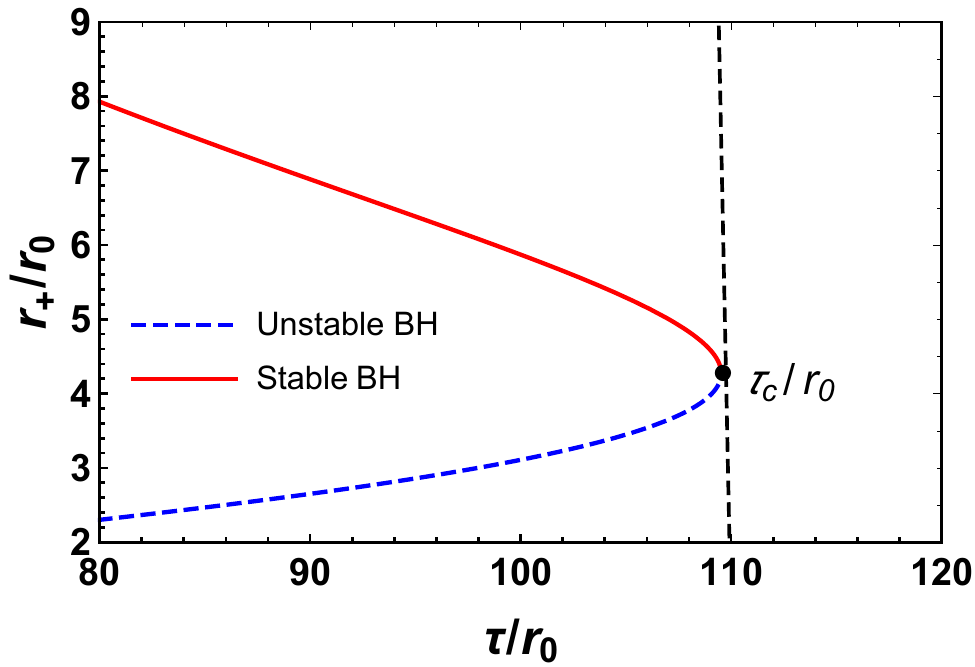} \hspace{-0.5cm}
\caption{ \textbf{Up panel:}Unit vector $n=\left(n^1, n^2\right)$ shown in $\Theta$ vs $r_{+} / r_0$ plane for $\tau / r_0=100, U/ r_0=0.1, C_{1}/ r_0=-0.1,C_{2}/ r_0=-0.3$ and $P r_0^2=0.001$. The black dots represent the zero points.\textbf{ Down panel:}The zero points of $\phi^{r+}$ in $\tau / r_0$ vs $r_{+} / r_0$ plane for black holes in GCE ($ U/ r_0=0.1, C_{1}/ r_0=-0.1,C_{2}/ r_0=-0.3$ and $P r_0^2=0.001$)}
		\label{nn4}
	\end{figure}
Now, let us embark on examining the topological structure of thermodynamics in AdS black holes. By considering Eqs. (\ref{ph2}), (\ref{pp1}) and (\ref{ph1}), we can derive the thermodynamic function
\begin{equation}
\Phi=\frac{1}{\sin \theta}\left(\frac{r_{+} C_1+2\left(1+U^2 \eta+C_2\right)}{4 \pi r_{+}}\right).
\end{equation}
The vector field $\phi$ consists of the following components
\begin{figure}
		\centering
        \includegraphics[scale = 0.3]{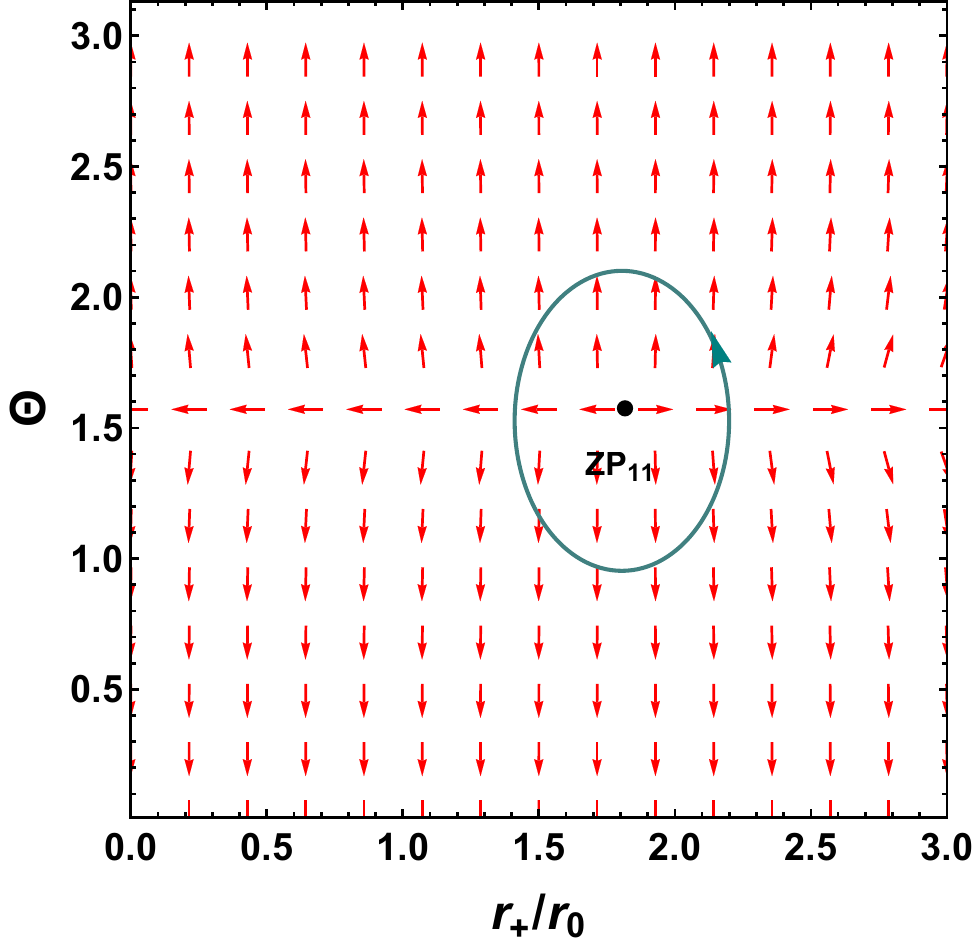} \hspace{-0.5cm}
		\includegraphics[scale = 0.3]{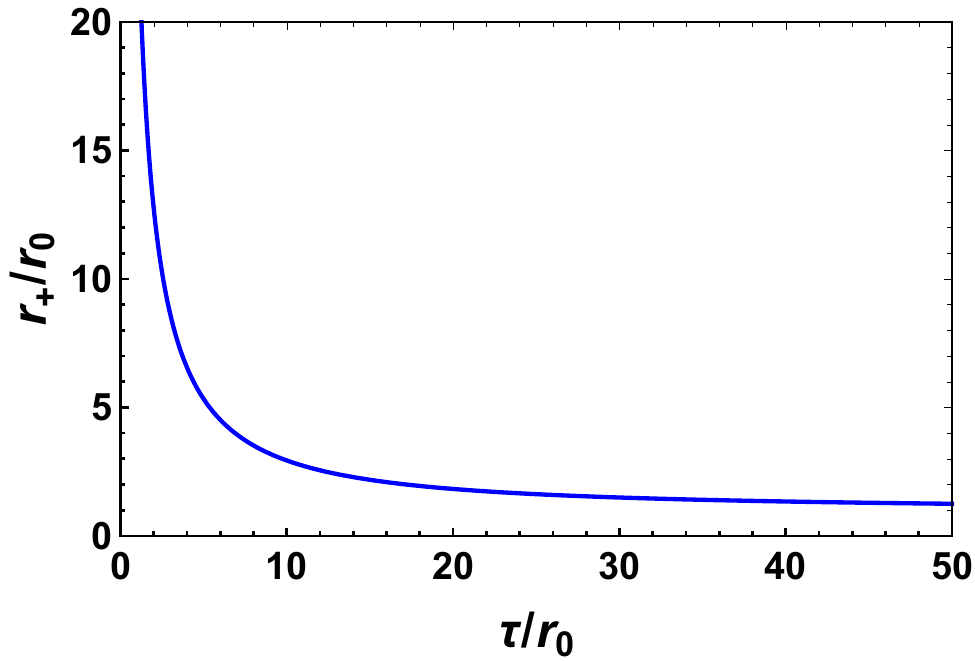} \hspace{-0.5cm}
\caption{ \textbf{Up panel:}Unit vector $n=\left(n^1, n^2\right)$ shown in $\Theta$ vs $r_{+} / r_0$ plane for $\tau / r_0=20,  C_{1}/ r_0=-0.1,C_{2}/ r_0=-0.3$ and $P r_0^2=0.02$. The black dots represent the zero points.\textbf{ Down panel:}The zero points of $\phi^{r+}$ in $\tau / r_0$ vs $r_{+} / r_0$ plane for black holes in GCE ($  C_{1}/ r_0=-0.1,C_{2}/ r_0=-0.3$ and $P r_0^2=0.02$)}
		\label{nn2}
	\end{figure}
\begin{equation}
\phi^{r+}=\frac{\csc \theta C_1}{4 \pi r_{+}}-\frac{\csc \theta\left(r_{+} C_1+2\left(1+U^2 \eta+C_2\right)\right)}{4 \pi r_{+}^2},
\end{equation}
and
\begin{equation}
\phi^\theta=-\frac{\cot \theta \csc \theta\left(r_{+} C_1+2\left(1+U^2 \eta+C_2\right)\right)}{4 \pi r_{+}}.
\end{equation}
After following the procedure outlined earlier, it was noted that there are no critical points in this system. Additionally, since the contour does not encompass any critical point, at $\vartheta=2 \pi$ (see Fig. \ref{f1}), the function $\Omega(\vartheta)$ attains a value of $0$.

Furthermore, let us view the solution of considered black holes as the topological thermodynamic defects. By employing the modified generalized free energy, we can derive
\begin{equation}
\begin{aligned}
\mathcal{F}&=E-\frac{S}{\tau}-Uq \\
&= \frac{r_{+}}{8 \pi}+\frac{U^2 \eta r_{+}}{8 \pi}-\frac{\tau U^2 r_{+}}{\tau}-\frac{r_{+}^2}{4 \tau}+\frac{P r_{+}^3}{3}\\
&+\frac{r_{+}^2 C_1}{16 \pi}+\frac{r_{+} C_2}{8 \pi}.
\end{aligned}
\end{equation}
The components of the vector field, as given by Eq. (\ref{wei10}), can be described as follows:
\begin{equation}
\phi^r=\frac{1}{8 \pi}+\frac{U^2 \eta}{8 \pi}-\frac{\tau U^2}{\tau}-\frac{r_{+}}{2 \tau}+P r_{+}^2+\frac{r_{+} C_1}{8 \pi}+\frac{C_2}{8 \pi},
\end{equation}
and
\begin{equation}
\phi^{\Theta}=-\cot \Theta \csc \Theta.
\end{equation}
By finding the solution to $\phi^{r}=0$, we can derive
\begin{equation}
\tau = \frac{4 \pi r_{+}}{1-8 \pi U^2+U^2 \eta+8 P \pi r_{+}^2+r_{+} C_1+C_2}.
\end{equation}
\begin{figure}
		\centering
        \includegraphics[scale = 0.3]{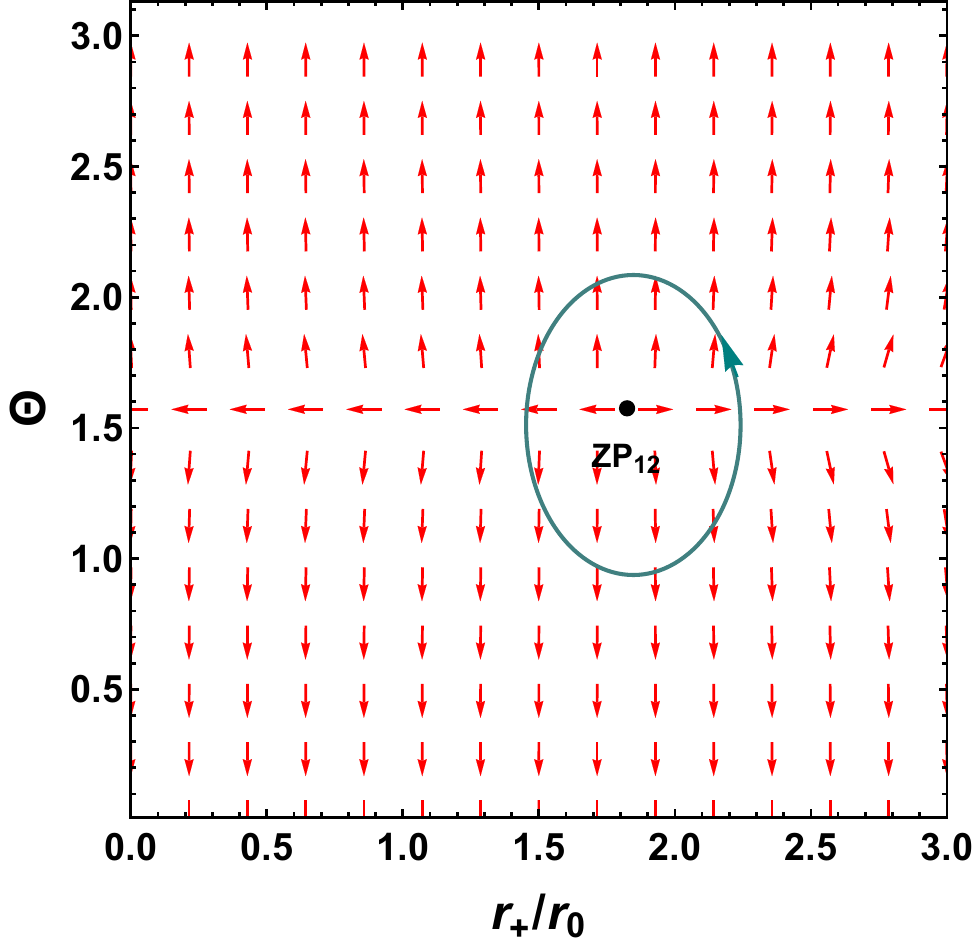} \hspace{-0.5cm}
		\includegraphics[scale = 0.3]{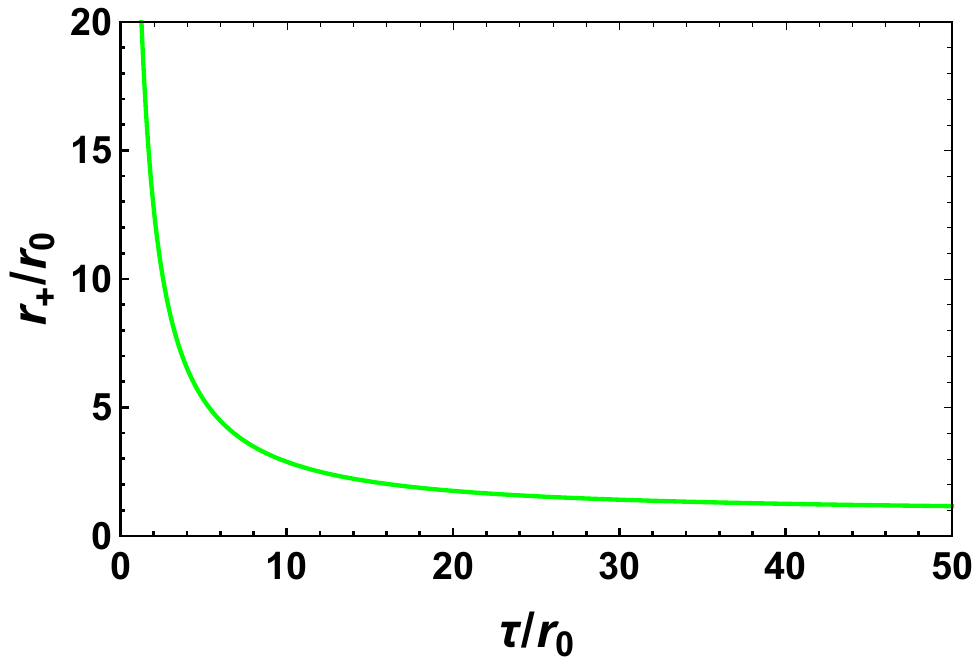} \hspace{-0.5cm}
\caption{ \textbf{Up panel:}Unit vector $n=\left(n^1, n^2\right)$ shown in $\Theta$ vs $r_{+} / r_0$ plane for $\tau / r_0=20,  C_{1}/ r_0=-0.1,C_{2}/ r_0=-0.3$ and $P r_0^2=0.02$. The black dots represent the zero points.\textbf{ Down panel:}The zero points of $\phi^{r+}$ in $\tau / r_0$ vs $r_{+} / r_0$ plane for black holes in GCE ($  C_{1}/ r_0=-0.1,C_{2}/ r_0=-0.3$ and $P r_0^2=0.02$)}
		\label{nn7}
	\end{figure}
Initially, we consider the Maxwell field to be a phantom with $\eta=-1$. It can be observed in Fig. \ref{nn1} that there are two zero points, $Z P_{7}$ $(2.84602, \pi / 2)$ and $Z P_{8}$ $(6.13285, \pi / 2)$, each having winding numbers of $-1$ and $+1$, respectively. As a result, the topological number is determined to be zero. Contrary to our findings in the canonical ensemble, Fig. \ref{nn1} exhibits two distinct sections for black holes: one for $\tau<\tau_c$ and another for $\tau>\tau_c$. The former section corresponds to an area where black holes are not stable (blue dotted line), while the latter section represents a region of stable black holes(red solid line) in Fig. \ref{nn4}. In the region of instability, the winding number of any zero point is $w=-1$, while in the stable region, it has a winding number of $w=+1$.  Therefore, the value of the topological number is $W=0$, which differs from what we observed in the canonical ensemble. The generation point is situated at $\tau / r_0=\tau_c / r_0=114.239$. Furthermore, in the CEM with $\eta=1$, the topological number is also $W=0$ and there are two zero points, namely $Z P_{9}$ at coordinates $(3.1092, \pi / 2)$ and $Z P_{10}$ at coordinates $(5.86967, \pi / 2)$. The  generation point is located at $\tau / r_0=\tau_c / r_0=109.526$. This indicates that the system's topological structure remains unaltered.

Interestingly, with an increase in the electric potential to $U=0.2$, a noteworthy observation arises: at the point $(1.8253, \pi / 2)$, when $\eta=-1$, a significant value of $Z P_{11}$ is noted. On the other hand, in the case of $\eta=1$, a single branch of the black hole is identified at coordinates $(1.75047, \pi / 2)$ with a winding number of $+1$ and a zero point. It can be inferred that within the system depicted in Figs. \ref{nn2} and \ref{nn7}, there are neither generation nor annihilation points present.

\section{Conclusions}\label{sec5}

In summary, this study explores the thermodynamic structure of phantom AdS black holes in the massive gravity in two different ensembles. Initially, we  employ Duan's $\varphi$-mapping theory  to calculate the topological charge linked with the critical point. Subsequently, we scrutinize the local and global topology of these black holes by assessing the number of cycles present at these defects. The primary key findings can be succinctly summarized as follows:

$(i)$ The canonical ensemble only supports validity for $\eta=1$ in classical Einstein-Maxwell theory, in the case of black holes possessing a total topological charge of negative one, there is only one conventional critical point ($CP_{1}$) present. Along an isobaric curve at this critical point, an increase in pressure leads to a decrease in phase number. This specific critical point ($CP_{1}$) marks the phase annihilation point. Black holes are considered as topological defects. The research indicates that black holes possess a total topological charge of $1$, while also revealing generation and annihilation points occurring at pressures below the critical pressure.

$(ii)$  The grand canonical ensemble maintains a constant electric potential in relation to the electric charge, without any critical point present in the system. If a topological defect representing a black hole exists within the thermodynamic space, we can observe either a generation point or absence of generation/annihilation point depending on the value of the electric potential. This value also determines the total topological charge, which can be either $0$ or $1$. In scenarios involving massive gravity, ensembles significantly influence the topological characteristics of phantom AdS black holes.

$(iii)$ The current study focuses on examining the impact of the canonical and grand canonical ensembles on black hole topological nature and stability. However, it is worth exploring other mixed ensembles \cite{l21} that could potentially represent the system's topological structure. Additionally, an analysis of the Joule-Thomson effect \cite{ch29} in the extended phase space is conducted to assess black hole stability. These aspects present intriguing avenues for further investigation.

\begin{acknowledgments}
We are greatly indebted to the two anonymous referees for their constructive comments to improve the presentation of this work. This work is supported by the Doctoral Foundation of Zunyi Normal University of China (BS [2022] 07, QJJ-[2022]-314), by the National Natural Science Foundation of China (NSFC) under Grants No. 12265007, No. 12205243 and No. 12375053, by the Sichuan Science and Technology Program under Grant No. 2023NSFSC1347,  by the Doctoral Research Initiation Project of China West Normal University under Grant No. 21E028, and by the Long-Term Conceptual Development of a University of Hradec Kr\'alov\'e for 2023, issued by the Ministry of Education, Youth, and Sports of the Czech Republic.
\end{acknowledgments}

\end{document}